\documentclass{amsart}

\usepackage{amsthm}
\usepackage{amsmath}
\usepackage{amsfonts}
\usepackage{amssymb}
\usepackage[T1]{fontenc}
\usepackage{yfonts}
\usepackage{bbold}
\usepackage[all]{xy}
\usepackage{dsfont}
\usepackage{upgreek}
\usepackage{mathrsfs}
\usepackage{enumerate}

\xyoption{all}

\theoremstyle{plain}

\newtheorem{thm}{Theorem}[section]
\newtheorem{prop}[thm]{Proposition}

\theoremstyle{definition}

\newtheorem{rmk}[thm]{Remark}

\newtheorem{ex}[thm]{Example}

\numberwithin{equation}{section}

\newcommand{\wt}{\widetilde}
\newcommand{\wh}{\widehat}
\newcommand{\mds}{\mathds}
\newcommand{\mfr}{\mathfrak}
\newcommand{\mcal}{\mathcal}

\newcommand{\ra}{\rightarrow}

\newcommand{\tra}{\twoheadrightarrow}

\title{The existence of a canonical lifting of even Poisson Structures to the Algebra of Densities}

\author{ A. Biggs }

\begin{document}

\begin{abstract} In this note we construct a canonical lifting of arbitrary Poisson structures on a manifold to its algbera of densities. Using this construction we proceed to classify all extensions of a fixed structure on the original manifold to its algebra of densities. The question is analogous to the problem studied by H.M.Khudaverdian and Th.Voronov for odd Poisson structures and differential operators. Although the questions are similar the results are distinctly marked, namely in the case of even Poisson structures there always exists a lift which is naturally defined. The proof of this result bears a remarkable resemblance to the construction of the Frolicher-Nijenhuis bracket. \end{abstract}

\maketitle

\section{Introduction}

Let $M$ be a supermanifold and let $C^{\infty}(\Pi T^{*} M)$ denote functions on the parity reversed cotangent bundle, $\Pi T^{*} M$. We have a natural subalgebra of $C^{\infty}(\Pi T^{*} M)$, $\mfr{A}(M)$, consisting of those functions that are polynomial in the fibre-coordinates and we shall denote the space of polynomial functions with degree $r$ by $\mfr{A}^{r}(M)$. Given an element $S \in C^{\infty}(\Pi T^{*}M)$ we can define an $r$-ary bracket on $M$ given by the formula, see \cite{Vor} for a detailed study:
\begin{equation} \{ f_{1} \cdots f_{r} \}_{S,r} := i^{*}( \cdots ((S,f_{1}),f_{2}) \cdots ,f_{r}), \end{equation}
where $($ , $)$ denotes the canonical odd symplectic structure on $\Pi T^{*}M$, $i:M \ra \Pi T^{*}M$ denotes the zero section, and we consider $C^{\infty}(M) \subset C^{\infty}(\Pi T^{*} M)$. If $S$ is an even function then we have that the induced brackets are anti-symmetric: $\{ f, g \} = (-)^{(\wt{f}+1)(\wt{g}+1)} \{g,f \}$.
 One can show, see \cite{Vor} or \cite{derbrack}, that any notion of an anti-symmetric bracket on $M$ can be derived in this way and we shall just take it as a definition. The master equation $(S,S) = 0$ translates to the Jacobi identity for $\mfr{A}^{2}(M)$, and an $L_{\infty}$-structure in general. Just a caveat to the reader to notice that if one considers the bracket $\{ \cdots \}_{S,r}$ as an operator then its parity is not equal to that of $S$ as a function. This is due to the oddness of the bracket and we see that the parity of the bracket $\{ \cdots \}_{S,r}$ is equal to $\wt{S} + r$.



The algebra of densities, denoted by $\mathcal{F}(M) = \oplus_{\lambda} \mathcal{F}^{\lambda}(M)$, is the algebra generated by densities of arbitrary weight, see \cite{KhVor1} or \cite{KhVor2} for more details. The multiplication is induced by the natural isomorphism $\mcal{F}^{\lambda}(M) \otimes \mcal{F}^{\mu}(M) \cong \mcal{F}^{\lambda+\mu}(M)$. It can be considered as an algebra generated by locally adjoining to $C^{\infty}(M)$ an invertible element denoted $t$. An element of this space, $s \in \mathcal{F}(M)$, is given in a local coordinate system by:
\[ s = \sum_{\lambda} s_{\lambda} t^{\lambda} , \]
where the sum is finite and $t$ transforms with the Jacobian: $\wt{t} = |J| t$. The new variable $t$ has even parity, is invertible and has weight 1 (all base coordinates have weight 0 in this picture). We shall denote by $\wh{M}$ the space whose algebra is $\mcal{F}(M)$. We have a natural notion of $T^{*}\wh{M}$, where we introduce an even momentum $p_{0}$, which transforms as $\partial_{t}$ and has weight -1. We can then define the space $\Pi T^{*} \wh{M}$, where we have the local coordinates $(x^{a},t;\wh{x}^{*}_{a},w^{*})$, where $w^{*} = t t^{*}$ is introduced becasue it has weight 0 and regularly crops up. We have used the notation $\wh{x}^{*}_{a}$ instead of $x^{*}_{a}$ as it transforms in a different way as we will see below. All algebraic data for ordinary manifolds can be extended to $\wh{M}$, in particular $\Pi T^{*} \wh{M}$ has a natural odd Poisson structure and hence we have the notion of anti-symmetric brackets on $\wh{M}$, in the text we shall give examples of what such brackets are like.

The main tool we shall use is that of the divergence map $\Delta: \mfr{A}^{\bullet}(M) \otimes \mathcal{F}^{1}(M) \ra \mfr{A}^{\bullet -1}(M) \otimes \mcal{F}^{1}(M)$. In local coordinates the operator is given by 
\[\Delta(s |Dx|) =  (-)^{\wt{a}}\partial_{x^{a}}\partial_{x^{*}_{a}}(s) |Dx|.\]
The divergence operator squares to zero and can clearly be extended to the full space $C^{\infty}(\Pi T^{*}M)\otimes \mcal{F}^{1}(M)$. Given a volume form $\rho$ on the manifold $M$, we can define the divergence with respect to $\rho$ of an element $S \in C^{\infty}(\Pi T^{*}M)\otimes \mcal{F}^{1}(M)$ by $\Delta_{\rho}(S) = \rho^{-1} \Delta(S\rho)$. For more details on the divergence operator and its relation to Batalin-Vilkovisky geometry see \cite{BV}, \cite{Kh3} and \cite{Kh1}. We have the following local equation:
\begin{equation}\label{eq_BV} \Delta_{|Dx|}(f,g) = - (\Delta_{|Dx|}f,g) - (-)^{\wt{f}}(f,\Delta_{|Dx|}g). \end{equation}
This equation says that $\Delta_{|Dx|}$ is a derviation of $($ $,$ $)$. The minus sign origintes from the oddness of both the bracket and the operator and the fact we only consider differential operators acting from the right.\footnote{Using the fact that $f\overleftarrow{\partial_{a}} = (-)^{\wt{a}(\wt{f} + 1)} \partial_{a} f$, a simple calculation shows that the odd bracket we use is $(-)^{\wt{f}}$ times the usual Batalin-Vilkovisky bracket. The reason why we use this one is because it satisfies very nice parity properties. We therefore have that the preceeding equation exactly translates to the usual derivation of the bracket by $\Delta$.}

\section{Lifting brackets to the algebra of densities}

We shall introduce a canonical odd second order differnetial operotor on $T^{*}\wh{M}$. This operator allows us to define a natural map $C^{\infty}(\Pi T^{*}M) \ra C^{\infty}(\Pi T^{*}\wh{M})$. Moreover this operation commutes with the odd bracket and hence allows us to classify all extensions of a given Poisson structure to the algebra of densities. 

\subsection{The canonical operator.} Recall that given a manifold $M$ with a volume form $\rho$ we have the operator $\Delta_{\rho}$, on the odd cotangent bundle, defined by the diagram (see \cite{Kh3}):

\[ \xymatrix{ C^{\infty}(\Pi T^{*} M ) \ar[r]^{ \times \rho} \ar@{.>}[d]^{\Delta_{\rho}} &C^{\infty}(\Pi T^{*}M) \otimes \mcal{F}^{1}(M) \ar[d]^{\Delta} \\ C^{\infty}(\Pi T^{*} M) & C^{\infty}(\Pi T^{*}M)\otimes \mcal{F}^{1}(M) \ar[l]^{ \times \rho^{-1} } } \]

The algebra of densities has a canonical volume form, $\rho_{0}$, which we in local coordinates has the form
\begin{equation}\label{volform} \rho_{0} = \frac{1}{t^{2}} |D(x,t)| .\end{equation}

Due to the existence of this volume form we have that the space $\wh{M}$ has a canonical odd second order operator, which we shall denote by $\delta$. In local coordinates we find an expression for $\delta$ of the form
\begin{equation} \delta = \Delta_{|Dx|} + \partial_{t^{*}} \left( \partial_{t} - \frac{2}{t} \right) \end{equation}

\begin{rmk} In two dimensions (on an ordinary orientated manifold) a volume form is equivalent to a symplectic form. This follows due to the fact that in this situation there is an isomorphism $\mcal{F}^{1}(M) \cong \Omega^{2}(M)$, and this will send a nowhere vanishing density to a nowhere vanishing two form which is a priori closed, i.e. symplectic. Therefore we see that on $S^{1}$ the space of densities has a natural symplectic structure. Although we shall not need this structure it has been used elsewhere, for example see \cite{Ov1} . \end{rmk}

\subsection{The canonical lift.}

Recall that the subalgbera $C^{\infty}(M) \subset \mcal{F}(M)$ consists of those functions with weight 0. As the odd bracket has weight 0, we have that $\wh{w}(f,g) = (\wh{w}(f), g) + (f,\wh{w}(g))$. Therefore if we take an element $S$ of weight 0 in $C^{\infty}(\Pi T^{*}\wh{M})$ we have that it induces a family of brackets on $M$ by using the standard formula. This is in fact part of a more general procedure, namely we have a well defined morphism of graded algebras, $C^{\infty}(\Pi T^{*} \wh{M}) \tra C^{\infty}(\Pi T^{*} M) \otimes \mcal{F}(M)$, which in coordinates corresponds to $t^{*} \mapsto 0$ (one can see surjectivity using a partition of unity argument for example). So, in general, by a lift for a fixed weight $\lambda$ we mean a section of this projection on the part of weight $\lambda$. In the case $\lambda = 0$ we see that this also corresponds to a lift of brackets.

\begin{thm} \label{t-1lift} Let $\lambda \neq 1$, then there exists a canonical map $C^{\infty}(\Pi T^{*}M) \otimes \mcal{F}^{\lambda}(M)  \rightarrow C^{\infty}( \Pi T^{*} \wh{M})$ that preserves weight, parity and degree on the subalgebra of polynomial functions. Moreover this map commutes with the odd bracket. The local form of this map is
\begin{equation} \xymatrix{ s(x,x^{*})|Dx|^{\lambda} \ar@{|->}[r] & \widehat{s} =  s(x,\widehat{x}^{*})t^{\lambda} + \frac{1}{1-\lambda}w^{*}t^{\lambda} \Delta_{|Dx|} s(x,x^{*})|_{x^{*} =\widehat{x}^{*}} } \end{equation} \end{thm}

The proof of this theorem proceeds as follows: Firstly fix an element $s \in C^{\infty}(\Pi T^{*}M) \otimes \mcal{F}^{\lambda}(M)$ and consider $p^{-1}(s) \subset C^{\infty}(\Pi T^{*} \wh{M})$, the set of those functions that are projected to $s$ by the above map. Then we find that for $\lambda \neq 1$ there exists a unique element in this space that satisfies the differential equation $\delta \wh{s} = 0$, and this defines the lift. The proof is therefore very similar to that of the construction of the Frolicher-Nijenhuis bracket where we can find a unique lift such that it is in the kernel of $\wh{d} = dx^{a}\partial_{a} - (-)^{\wt{a}} x^{*}_{a}\partial_{dx^{*}_{a}}$.

\begin{proof} Looking at an arbitrary element $S_{0} \in C^{\infty}(\Pi T^{*}M)\otimes \mcal{F}^{\lambda}(M)$, we see that a general element that lies above it, of the type required in the theorem, is of the form $t^{\lambda}S_{0}(x,\wh{x}^{*}) + t^{*}t^{\lambda +1}S_{1}(x,\wh{x}^{*})$. We apply $\delta$, the canonical operator defined in the last section, to this lift:
\[ \delta( t^{\lambda}( S_{0} + t^{*}t S_{1}) ) = t^{\lambda} \left( \Delta_{|Dx|} (S_{0}) + (\lambda - 1) S_{1} \right) - t^{*} t^{\lambda+1}\Delta_{|Dx|}(S_{1}) .\] 
We see that for $\lambda \neq 1$ we have that $S_{1}$ is uniquely defined given an $S_{0}$, if we require that this is in the kernel of $\delta$, which we glean as being:
\[ S_{1} = \frac{1}{1-\lambda} \Delta_{|Dx|}S_{0} .\]

We see that this gives the lift as claimed. As a double check to see this is well defined we could explicitely check how the coordinates $\widehat{x}^{*}_{a}$
transform, i.e. as parity reversed momenta:
\[ \widehat{x}^{*}_{a'} = \partial_{a'} x^{a} \widehat{x}^{*}_{a} +
(-)^{\widetilde{b'}}\partial_{a'b'}x^{a}\partial_{a}x^{b'}w^{*} .\]
Using this we can easily check that this lift is well defined, i.e. coordinate independant. We now need to check that this commutes with the odd bracket, i.e. $(\widehat{s} , \widehat{u} ) = (s,u)^{\wedge}$. We find
\[ (\widehat{s} , \widehat{u} ) = (s,u) - w^{*}\left[ (\Delta s,u) + (-)^{\widetilde{s}}(s,\Delta u) \right] \]
Therefore that the lift commutes with the bracket iff $\Delta_{|Dx|}(s,u) = - (\Delta_{|Dx|} s, u) - (-)^{\widetilde{s}} (s, \Delta_{|Dx|} u)$. But this is just equation~\eqref{eq_BV}, and hence the result follows.\end{proof}

We now study the geometric data contained in this map. As for the long bracket in the symmetric case we see that the extension of an $r$-ary bracket requires the following data $\{ x^{a_{1}} , \cdots x^{a_{r-1}} , t \} = T^{a_{1} \cdots a_{r-1}} t$, and this $T$ transforms us an upper connection over the tensor $S^{a_{1} \cdots a_{r}} = \{x^{a_{1}}, \cdots ,x^{a_{r}} \}$, considered as a map $\Omega^{1}(M) \ra \mfr{A}^{r-1}(M)$. The map above states that there exists a canonical upper-connection for a $\Pi$-symmetric tensor $S$ and this is given by $T = \Delta_{|Dx|} S$. 

We shall now give some examples of this map.

\begin{ex} Let $X = X^{a}x^{*}_{a}$ be a parity reversed vector field on $M$. Then the lift of $X$ is $ \widehat{X} = X^{a}\widehat{x}^{*}_{a} + (-)^{\widetilde{a}(\widetilde{X} +1)}\partial_{a}X^{a}w^{*}$. This is of course just the parity reversed Lie derivative, $\Pi \mathcal{L}_{\Pi X}$. \end{ex}

\begin{ex} Let $\pi = \pi^{ab} x^{*}_{a}x^{*}_{b}$ be a 2-ary bracket on $M$. The lift of $\pi$ has the form:

\begin{equation} \widehat{\pi} = \pi^{ab}\widehat{x}^{*}_{a}\widehat{x}^{*}_{b} + 2 (-)^{ \widetilde{\pi}\widetilde{a}} \partial_{a} \pi^{ba}\widehat{x}^{*}_{b} w^{*} \end{equation}



If our Poisson bracket comes from a symplectic structure then in canonical coordinates the lift is trivial, as would be expected for we have a natural volume form.

\end{ex}

\begin{ex} In this example we shall explore the case of a (super) Lie algebra $\mathfrak{g}$. The Lie algebra structure on $\mathfrak{g}$ induces a Poisson structure on $\mathfrak{g}^{*}$ and thus on the space of densities on $\mathfrak{g}^{*}$. In this example it is in fact more natural to consider the algebra that is polynomial in the variable $t$ rather than letting it be invertible. Then the space associated to the algebra of densities, $\widehat{\mathfrak{g}}^{*}$, is isomorphic to $\mathfrak{g}^{*}\times \mathds{K}$. This induces a Lie algebra structure on $\widehat{\mathfrak{g}}$, as the Poisson bracket is of the correct weight, and therefore we have the following sequence of Lie algebras

\[ \xymatrix{ 0 \ar[r] & \mathfrak{g} \ar[r] & \widehat{\mathfrak{g}} } \]

This can be extended to a sequence of $\mathfrak{g}$-modules by taking the quotient of the extended Lie algebra by $\mathfrak{g}$. The final term in the sequence is a one dimensional vector space, and hence isomorphic $\mathds{K}$. Recall that extensions of Lie algebras, as above, is equivalent to an action of $\mfr{g}$ on $\mds{K}$, we shall now see what this action looks like.

Let us pick a basis on $\mathfrak{g}$, say $\{ x^{i} \}$ with structure constants $C^{ij}\textrm{}_{k}$. This basis then allows to make an identification  $\widehat{\mathfrak{g}}^{*} \cong \mathfrak{g}^{*} \times \mathds{K}$. The Poisson bracket and its lift have the form

\[ \pi = \frac{1}{2} C^{ij}\textrm{}_{k}x^{k} x^{*}_{j}x^{*}_{i} \]

\[ \widehat{\pi} = \frac{1}{2} C^{ij}\textrm{}_{k}x^{k} x^{*}_{j}x^{*}_{i} + w^{*} (-)^{\widetilde{i}}C^{ji}\textrm{}_{i}x^{*}_{j} \]

Thus the action on $\mathds{K}$ has the from, $\mathfrak{g} \rightarrow \mathfrak{gl}(1)$, given by $x \mapsto \textrm{str}(x)$. A simple check shows that this is invariant of our choice of basis. 

\end{ex}

\subsection{Lifting fixed Poisson structures}

Now we fix a manifold $M$ and an element $s \in \mfr{A}^{r}(M)$ such that $(s,s) = 0$. Using the above map we shall classify the space of all $S$ that lie above $s$, satisfy $(S,S) = 0$, as well as having the same weight and degree as $s$. Firstly consider $S - \widehat{s}$, then as this induces the zero bracket on functions it follows that $S - \widehat{s} = Q w^{*}$, and, by the restrictions on degree, $Q$ is a polynomial of $r-1$ in the fibre coordinates, and has parity (as a function) opposite to that of $s$. If we locally expand $Q$ we see that any term containing $w^{*}$ is killed by the $w^{*}$ multiplying $Q$ and hence we have that

\[ S - \wh{s} = Q^{a_{1} \cdots a_{r-1}}\wh{x}^{*}_{a_{1}} \cdots \wh{x}^{*}_{a_{r-1}} w^{*}. \]

Checking the coordinate transformation we see that $Q^{a_{1} \cdots a_{r-1}} x^{*}_{a_{1}} \cdots x^{*}_{a_{r-1}}$ is a well defined element of $\mfr{A}^{r-1}(M)$. We therefore see that for a general $s \in \mfr{A}^{r}(M)$ the space of $S \in \mfr{A}^{r}(\wh{M})$ that restrict to $s$ is an affine space modelled on $(\mfr{A}^{r-1}(M))^{\wt{s} +1}$.

 We now enforce the condition that $(S,S) = 0$. To do this we shall need the following properties of the odd bracket $($ $,$ $)$:

\[ (fg,h)= (-)^{\widetilde{f}}f(g,h) + (-)^{\widetilde{g}\widetilde{h}}(f,h)g \]

\[ (f,g) = (-)^{\wt{f}\wt{g}}(g,f) \]

Then writing $S = \wh{s} + Qw^{*}$ we find that

\[ (S,S) = (\wh{s} + Qw^{*} , \wh{s} + Qw^{*}) = \left(1 + (-)^{\wt{s} } \right)(s,Q)w^{*} \]

We sum these calculations up in the following proposition.

\begin{prop} Let $s$ be a Poisson r-ary bracket, i.e. $(s,s)=0$ and $\wt{s} = 0$. Then the space of brackets above $s$ that are also Poisson and have the same degree, weight and parity, is isomorphic to the odd elements in $\textrm{Ker}(s, \cdot) \cap (\mfr{A}^{r-1}_{M})$. 
\end{prop}

\begin{ex} Let us take the standard symplectic structure on $\mathds{R}^{2n}$, $\omega = p_{*}^{i}x^{*}_{i}$, then $\widetilde{\omega} = 0$. A vector field on $X$ is symplectic iff $(\omega,\Pi X) =0$. This is the case if $X^{i}\textrm{}_{,j} =  - X_{j}\textrm{}^{,i}$, $X^{i,j} = X^{j,i}$, and $X_{i,j} = X_{j,i}$. From the above proposition we see that the space of lifts  of $\omega$ is classified exactly by the space of symplectic vector fields. Let us take a symplectic vector field $X$, we then have the lift $\Omega_{X} = \wh{\omega} + \Pi X w^{*}$. The extra term in the bracket is given by:

\[ \{ H ,t \} = ((\Omega_{X},H),t) = - X(H) t .\]

On $\mds{R}^{2n}$ every symplectic vector field is Hamiltonian and hence there exists a function $f$ such that $X = X_{f}$ and hence we may rewrite the above as $\{ H,t\} = \{H,f\} |Dx|$. Let us consider in detail how this defines a time evolution of $\lambda$-densities. Fix a Hamiltonian $H$ on $M$ and take $\Psi = \psi t^{\lambda}$. We find that the time evolution of this density is given by

\[ \frac{d}{d\tau} \Psi(\tau) = ((\Omega_{X},H),\Psi) = \left( \{ H,\psi\} + \lambda \{H,f\} \psi \right)t^{\lambda}  = (\dot{\psi} + \lambda \dot{f} \psi )t^{\lambda} \]

\end{ex}

\begin{ex}{Q-manifolds.} A $Q$-manifold is a pair, $(M,Q)$, where $M$ is a manifold and $Q$ is an odd derivation, also known as a homological vector field, on $M$ that squares to zero, $Q^{2} = \frac{1}{2}[Q,Q] = 0$, for more details see \cite{V1} or . In terms of the $\Pi T^{*} M$ an odd vector field is nothing other than an even element of $\mfr{A}^{1}(M) \ni Q$ and the fact that it squares to zero is just $(Q,Q) =0$. We can thus classify all extensions of $Q$ to the algebra of densities as above: The space of homological vector fields extending $Q$ is isomorphic to the kernel of $Q$.

For a specific example consider $\Pi TM$ with the vector field $d = dx^{a} \partial_{x^{a}}$.  Then $\wh{d} = dx^{a} \wh{x}^{*}_{a}$, the space $\Pi TM$ has a canonical volume form. Hence a general lift of this vector field is $dx^{a} \wh{x}^{*}_{a} + w^{*} \theta$, where $\wt{\theta} = 1$ and $d \theta = 0$, that is $\theta$ is a closed differential form of odd weight. \end{ex}

\section{Conclusions and Further Questions}

We have seen that on an arbitrary manifold, $M$, there exists a natural lift of any weighted function on the odd cotangent space, $C^{\infty}(\Pi T^{*}M) \otimes \mcal{F}^{\lambda}(M)$, to $C^{\infty}(\Pi T^{*}\wh{M})$ if $\lambda \neq 1$. Moreover we have also shown the existence of a natural differential operator on $\wh{M}$ that generates the odd bracket, $\delta$. There are however still a variety of questions that may still be of interest which we shall attempt to address at a later date, and we shall state some of them now:

\begin{enumerate}
\item The singular point of the lift, $\lambda = 1$, should be explained in greater detail. We have the natural isomorphism:

\[ C^{\infty}(\Pi T^{*}M) \otimes \mcal{F}^{\lambda}(M) \cong \mcal{F}^{\lambda /2}(\Pi T^{*} M) .\]

We therefore see that the singular point corresponds to half densities on the odd cotangent bundle, however this is still not a completely satisfactory description.

\item Of course the quantum version of all the above still remains to be studied. In particular, due to the existence of $\delta$, it is natural to study the quantum master equation on $\wh{M}$:

\[  \delta S + \frac{i}{2\hbar} (S,S) = 0.\]

\end{enumerate}

\section*{Acknowledgements}

I would like to thank both H. M. Khudaverdian and Th. Voronov for many enlightining discussions.

\end{document}